\documentstyle[twoside,fleqn,espcrc2]{article}

\newcommand{\AmS}{{\protect\the\textfont2
  A\kern-.1667em\lower.5ex\hbox{M}\kern-.125emS}}

\newcommand{\nc}{\newcommand}
\nc{\mlgraph}{{multiple-line graph }}
\nc{\mlgraphs}{{mul\-tiple\--line graphs }}
\nc{\Mlgraph}{{Multiple-line graph }}

\nc{\mldiagram}{{multiple-line diagram }}
\nc{\mldiagrams}{{multiple-line diagrams }}

\nc{\mlmoment}{{multiple-line moment}}
\nc{\mlmoments}{{multiple-line moments}}

\nc{\be}{\begin{equation}}
\nc{\ee}{\end{equation}}
\nc{\bea}{\begin{eqnarray}}
\nc{\eea}{\end{eqnarray}}
\nc{\bela}{\begin{eqnarray*}}
\nc{\eela}{\end{eqnarray*}}

\nc{\eqn}[1]{{(\ref{#1})}}
\nc{\cA}{{\cal A}}
\nc{\cB}{{\cal B}}
\nc{\cC}{{\cal C}}
\nc{\cD}{{\cal D}}
\nc{\cE}{{\cal E}}
\nc{\cF}{{\cal F}}
\nc{\cG}{{\cal G}}
\nc{\cH}{{\cal H}}
\nc{\cI}{{\cal I}}
\nc{\cJ}{{\cal J}}
\nc{\cK}{{\cal K}}
\nc{\cL}{{\cal L}}
\nc{\cM}{{\cal M}}
\nc{\cN}{{\cal N}}
\nc{\cO}{{\cal O}}
\nc{\cP}{{\cal P}}
\nc{\cQ}{{\cal Q}}
\nc{\cR}{{\cal R}}
\nc{\cS}{{\cal S}}
\nc{\cT}{{\cal T}}
\nc{\cU}{{\cal U}}
\nc{\cV}{{\cal V}}
\nc{\cW}{{\cal W}}
\nc{\cX}{{\cal X}}
\nc{\cY}{{\cal Y}}
\nc{\cZ}{{\cal Z}}

\nc{\simo}[1]{{\stackrel{#1}{\simeq}}}
\nc{\geqo}[1]{{\stackrel{#1}{\geq}}}
\nc{\geo}[1]{{\stackrel{#1}{>}}}
\nc{\guo}[1]{{\stackrel{#1}{\succ}}}

\nc{\rbo}{\raisebox}
\nc{\RR} {\rangle \! \rangle}
\nc{\LL} {\langle \! \langle}
\nc{\rmi}[1]{{\mbox{\small #1}}}
\nc{\eq}{eq.~}
\nc{\nr}[1]{(\ref{#1})}
\nc{\ul}{\underline}
\nc{\mc}{\multicolumn}
\nc{\todo}[1]{\par\noindent{\bf $\rightarrow$ #1}}

\nc{\cu}{{\cal u}}
\nc{\kprime}{{\,\prime}}

\title{Dynamical linked cluster expansions: Algorithmic aspects and
       applications}

\author{H.~Meyer-Ortmanns
        \thanks{e-mail: ortmanns@theorie.physik.uni-wuppertal.de }
        \address{Institute of Theoretical Physics,
        Bergische Universit\"at Wuppertal, 
        \\
        Gauss-Str.20, D-42097 Wuppertal, Germany}%
        and 
        T.~Reisz\thanks{e-mail: reisz@thphys.uni-heidelberg.de, 
        supported by a Heisenberg fellowship}
        \address{Institute of Theoretical Physics, University of 
         Heidelberg, \\
        Philosophenweg 16, D-69120 Heidelberg, Germany}}

\begin{document}

\begin{abstract}
Dynamical linked cluster expansions are linked cluster expansions with
hopping parameter terms endowed with their own dynamics. They amount
to a generalization of series expansions from 2-point to point-link-point
interactions. We outline an associated multiple-line graph theory involving
extended notions of connectivity and indicate an algorithmic implementation
of graphs. Fields of applications are SU(N) gauge Higgs systems within
variational estimates, spin glasses and partially annealed neural networks.
We present results for the critical line in an SU(2) gauge Higgs model for
the electroweak phase transition. The results agree well with corresponding
high precision Monte Carlo results.
\end{abstract}

\maketitle

\section{INTRODUCTION}

Linked cluster expansions (LCEs) provide an analytic alternative to 
large scale
Monte Carlo simulations. They are convergent series expansions of
thermodynamic quantities. 
Originally developed in the infinite volume, they
have recently been generalized to the finite volume \cite{hilde1} so that
the expansion allows for a finite size scaling analysis. If the expansion is
performed to a high order in the hopping parameter, even the critical
region becomes available from the symmetric phase in spite of the ansatz
that the expansion is performed about an ultralocal decoupled system
\cite{thomas1}.

\section{DLCEs}

For linked cluster expansions the action is typically split according
to
\be\label{act}
S(\phi,v)  = \sum_{x \in \Lambda}S^\circ(\phi_x)
             - \frac{1}{2} \, \sum_{x,y \in \Lambda} 
               v_{x,y}\phi_x\phi_y .
\ee
Here $S^\circ$ denotes the ultralocal part of the action depending on
generic fields $\Phi$ associated with single sites $x$ of a lattice 
$\Lambda$. The second term represents a pair interaction between fields
at points $x,y \in \Lambda$. A frequent choice for $v_{xy}$ is a coupling
between nearest neighbours of $\Lambda$, $v_{xy}$ is proportional to
the so called hopping parameter $\kappa$.

For dynamical linked cluster expansions the action is split into 2 ultralocal
parts $S^\circ(\Phi_x)$ and $S^1(U_l)$ with $S^\circ(\Phi_x)$ as
above and $S^1(U_l)$ a single link action, depending on a field $U_l$
associated with link $l=(x,y)$ of the lattice, 
and a point-link-point interaction,
depending on the hopping parameter $\kappa$ implicitly via $v_{xy}$.
It reads
\bea\label{2act}
S(\phi,U,v) & = & \sum_{x \in \Lambda}S^\circ(\phi_x)
             +\sum_{links \, l}S^1(U_l)
            \nonumber \\
        && - \frac{1}{2} \, \sum_{x,y \in \Lambda}
               v_{xy}\phi_xU_{x,y}\phi_y .
\eea
Note that the former constant parameter $v_{xy}$ gets replaced by 
$v_{xy}U_{xy}$ with $U_{xy}$ obeying a dynamics governed by $S^1$.
The action of (\ref{2act}) includes SU(N) gauge
Higgs systems within variational estimates, spin glasses, partially
annealed neural networks, but also 
generalizes to QCD with dynamical fermions if a
suitable variational ansatz is made.
From a systematic point of view, the generalization from 2-point
interactions in a hopping term of an LCE to a point-link-point
interaction in the hopping term of the class of actions in (\ref{2act})
amounts to a first step in the generalization to series expansions for
n-point interactions with $n>2$.
The graphical expansion for DLCEs arises from a Taylor expansion of
$\ln Z(H_x,I_l,v)$
about $v=0$ with external sources $H$, $I$.

In DLCEs a new type of connectivity arises. While formerly vertices are
connected via lines, here, in addition, lines are connected via a 
new type of vertices, represented by a dashed line, with
\be
  \left. 
{
\setlength{\unitlength}{0.8cm}
%
%
\begin{picture}(4.0,1.0)

%

\put(0.0,-0.6){
\setlength{\unitlength}{0.8cm}
\begin{picture}(4.0,0.0)

\qbezier(1.0,1.5)(2.0,2.1)(3.0,1.5)
\qbezier(1.0,1.0)(2.0,1.6)(3.0,1.0)
\qbezier(1.0,0.0)(2.0,0.6)(3.0,0.0)

\put(1.6,1.1){\makebox(0.2,0){$\cdot$}}
\put(1.6,0.8){\makebox(0.2,0){$\cdot$}}
\put(1.6,0.5){\makebox(0.2,0){$\cdot$}}

\put(2.0,1.85){\line(0,1){0.2}}
\put(2.0,1.6){\line(0,1){0.2}}
\put(2.0,1.35){\line(0,1){0.2}}
\put(2.0,1.1){\line(0,1){0.2}}
\put(2.0,0.85){\line(0,1){0.2}}
\put(2.0,0.6){\line(0,1){0.2}}
\put(2.0,0.35){\line(0,1){0.2}}
\put(2.0,0.1){\line(0,1){0.2}}

\end{picture}
}
%

\end{picture}
}
   \right\} \nu
  = 
   \left( \frac{\partial^\nu W(I)}{\partial I^\nu_l} \right)_{I=v=0}
\ee
and $W=\ln{Z}$.
Graphical expansions for correlation functions are generated from an
expansion of $W$ by attaching external $\Phi$-lines and $U$-lines.

We remark that the usual LCE is included in a DLCE for frozen
U-dynamics. Formally all multiple lines with $n>1$ are absent because
$\partial^nW(I)/\partial^n I_l=0$ for $n>1$.

{\it Multiple-line (m-line) graph theory}.
A complete multiple-line graph theory has been formulated in
\cite{hilde2}.
Here we only summarize the most important notions out of a glossary
which can be found in \cite{hilde2}.

A multiple-line graph is a structure
\be
  \Gamma = (\cL_\Gamma, \cM_\Gamma, \cB_\Gamma,
             E_\Gamma^{(\phi)}, E_\Gamma^{(U)},
             \Phi_\Gamma, \Psi_\Gamma).
\ee
in which ${\cal L}_\Gamma$, ${\cal M}_\Gamma$, ${\cal B}_\Gamma$
are three mutually disjoint
sets of bare internal lines, multiple-lines, and vertices of $\Gamma$,
respectively. $E_\Gamma^{(\Phi)}$ and $E_\Gamma^{(U)}$ are maps that
assign to every vertex $v$, multiple-line $m$ the number of external 
$\Phi$-lines, external $U$-lines to $v$, $m$, respectively.
Finally, $\Phi_\Gamma$ and $\Psi_\Gamma$ are incidence relations that
assign bare internal lines to their endpoint vertices and to their 
multiple-lines, respectively.

The central notion of {\it multiple-line connectivity} can be traced back to
the familiar notion of connectivity of LCE graphs by introducing auxiliary
graphs. {\it Topological equivalence} of multiple line graphs is
formulated  by means of incidence relations. Certain operations
on graphs must be defined such as the removal, the decomposition of an
m-line or the decomposition of a vertex. To characterize equivalence
classes of graphs, we define notions like
${\bf 1LI}$ (1-line irreducible), 
${\bf 1VI}$  (1-vertex irreducible), and
${\bf 1MLI}$  (1-multiple-line irreducible).
The definitions as well as further details can be found in \cite{hilde2}.

{\it Algorithmic aspects}.
An algorithmic construction of the graphs which contribute to the
n-point susceptibilities is the first step for a fully automatized
computer implementation of the generation of graphs. In the following
we specify a possible set of modules in the generation of DLCE-graph
classes. We start with the class 

${{\cal P}_1(L)}$: the set of connected 1LI LCE vacuum graphs
with $l$ internal lines. Apart from certain exceptions 
this set contains graphs 
whose vertices are connected by just one line. 
For the precise
definition we refer to \cite{thomas1}. The next module is

${{\cal P}_2(L)}$:  the set of connected 1LI LCE vacuum graphs with
$L$ internal lines whose vertices may be connected by any number
of bare lines. In the usual LCE one would attach external lines in the next
steps. In a DLCE we have first to generalize the set of LCE vacuum
graphs to include graphs with dashed lines and resolved vertices. This
way we are led to the classes

${\widetilde{{\cal MP}_2}(L)}$: the set of 1LI m-line
vacuum graphs with
$L$ internal bare lines so that all bare lines of an m-line have
the same endpoint vertices and

${{\cal MP}_2(L)}$: the set of all 1LI m-line vacuum graphs with $L$
internal lines. In particular this class contains graphs that would be
disconnected without dashed lines.
In the next modules we attach external $\Phi$-lines and $U$-lines
in specific ways and obtain

${{\cal Q}_k(L)}$: the set of renormalized vertex moment graphs with
$k$ external $\Phi$-lines and $L$ internal lines that are 1LI and have just
one external vertex and no external m-line.

${{\cal R}_k(L)}$: the set of renormalized m-line moment graphs with
$k$ external $U$-lines and $L$ internal lines that are 1LI and have
just one external m-line, but no external vertex.

${{\cal S}_{E_1,E_2}(L)}$: the set of graphs with $E_1$ external
$\Phi$-lines, $E_2$ external $U$-lines and $L$ internal lines
that are both 1VI and 1MLI. Because of certain factorization properties
of the corresponding analytic expressions, graphs of
${\cal Q}_k(L)$ ${\cal R}_k(L)$, and ${\cal S}_{E_1,E_2}(L)$ can
be patched together to yield the full set of connected 1LI DLCE graphs.

The additional type of connectivity leads to a pronounced proliferation
of DLCE graphs as compared to the LCE graphs for the same order in
the number of internal lines. As an example let us consider the 2-point
susceptibility $\chi_2$. The ratio of the number of DLCE to LCE graphs
contributing to $\chi_2$ 
to $O(\kappa^4)$ is already $80/8$.
Once the algorithmic generation of DLCE graphs is implemented on a
computer, the critical region of a variety of systems will become available
in spite of the ponounced proliferation of graphs.
 
\section{Applications}

{\it The SU(2) Higgs model.}
The action of the SU(2) Higgs model in 4 dimensions does not quite
have the form to allow for a DLCE, because the Wilson part of the action
is neither ultralocal nor of the type point-link-point.
Therefore we have replaced the action by a variational ansatz that leads
to a factorization of the partition function over time slices, whereas the
action for the degrees of freedom along the 3-d spacelike hypersurfaces
takes a form that is included as a special case of Eq. (\ref{2act}). 
The precise form
can be found in \cite{hilde2}. The equations that determine the optimal
choice of variational parameters are solved by means of a DLCE 
in 3 dimensions.
Here we directly state one particular result for the critical line 
of the electroweak
transition in the SU(2) Higgs model.
It refers to $\kappa_{crit}$
at $\lambda=5.0\cdot 10^{-4}$ and $\beta=8.0$.

\vskip12pt

\begin{tabular}{rll}
\hline
  Method & lattice & $\kappa_{crit}$
  \\ \hline
  Monte Carlo & $2\cdot 32\cdot 32 \cdot 256$ & $0.12887(1)$
    \\
  DLCE & $4\cdot \infty^3$ & $0.1282(1)$
    \\ \hline
\end{tabular}

\vskip12pt

The Monte Carlo data are taken from Fodor et al. \cite{fodor}.
``DLCE'' refers to variational estimates in combination with DLCEs
in which expectation values are evaluated including terms 
of $O(\kappa^4)$.
Such a moderate order in the expansion may hide the complexity of the 
analysis \cite{hilde3}. The total number of graphs contributing 
to the variational
equations is about several hundred. Thus the good quantitative agreement
with the high precision Monte Carlo results of Fodor et al. is not so
surprising. In addition we have been lucky in the choice of variational
ansatz. The 3-dimensional spacelike hypersurfaces are expected to contain
the nonperturbative degrees of freedom that drive the Higgs transition,
whereas the mean field treatment of the timelike degrees of freedom
seems to be a reasonable approximation, although the temperature
dependence gets lost this way by construction. 

{\it Avoiding the replica trick in spin glasses}.
The quantity that is of actual interest in spin glasses is 
$[[ln{Z_\beta(U)}]]$ in which
$Z_\beta$ denotes the partition function of a spin system at inverse
temperature $\beta$ for given interactions (couplings) $U$. $[[\cdots]]$
denotes an average over $U$ w.r.t. a probability distribution
governed by an
action of the type of $S^1$, cf. (\ref{2act}).
Usually the replica trick
involves an uncontrolled extrapolation from integer to real numbers.
In contrast to that we have shown in \cite{hilde2} that
$[[\ln{Z_\beta(U)}]]$ 
can be directly evaluated in a modified type of DLCE.
As a series representation
we have
\be \label{sg.wseries}
    [[ \frac{1}{V} \; \ln Z_{\beta}(U) ]] \; = \;
    \sum_{L\geq 0} (2K)^L 
     \sum_{\Delta} w(\Delta) .
\ee
The sum runs over all graphs $\Delta$ that belong to 
$\widetilde{{\cal G}}_{0,0}(L)$, the subset of multiple-line 
vacuum graphs
with $L$ internal lines that stay
connected without any dashed lines in the usual LCE sense. 
In (\ref{sg.wseries})
$w(\Delta)$ denotes the full analytic weight of the graph $\Delta$,
its explicit expression can be found in \cite{hilde2}.

We expect that the DLCE series in (\ref{sg.wseries}) is convergent for
a large class of interactions for which the set
$\widetilde{{\cal G}_{0,0}}(L)$
can be further restricted to certain subclasses.

\end{document}